\documentclass[12pt,a4paper]{article}%
\pdfoutput=1
\usepackage{amsmath,amssymb,amsfonts,wasysym}
\usepackage[pdftex]{graphicx}
\usepackage{color}
\usepackage{float}
\usepackage[bf,footnotesize]{caption2}
\usepackage[]{hyperref}%
\numberwithin{equation}{section}
\setcaptionmargin{1cm}

\def\La{\mathcal{L}}
\def\Amp{\mathcal{A}}
\def\({\left(}
\def\){\right)}
\def\f{\frac}
\def\be{\begin{equation}}
\def\ee{\end{equation}}

\def\de{\partial}
\def\demub{\de_{\mu}}
\def\denub{\de_{\nu}}

\def\demua{\de^{\mu}}

\begin{document}
\begin{titlepage}
\begin{flushright}
IFUP-TH/2010-17
\end{flushright}
\vskip 1.0cm
\begin{center}
{\Large \bf A ``composite'' scalar-vector system at the LHC} \vskip 1.0cm
{\large A. E. C\'arcamo Hern\'andez$^{a,b}$ and R. Torre$^{b,c}$}\\[0.7cm]
{\it $^a$ Scuola Normale Superiore, Piazza dei Cavalieri 7, I-56126 Pisa, Italy}\\[5mm]
{\it $^b$ INFN, Sezione di Pisa, Largo Fibonacci 3, I-56127 Pisa, Italy}\\[5mm]
{\it $^c$
Universit\`a di Pisa, Dipartimento di Fisica, Largo Fibonacci 3, I-56127 Pisa, Italy}
\end{center}
\vskip 1.0cm
\begin{abstract} 
In the framework of a strong dynamics for Electro-Weak Symmetry Breaking (EWSB), both vector and scalar degrees of freedom have been studied in the literature within an effective Lagrangian approach. Here we consider
the case in which both a scalar, $h$, and a vector, $V$ -- respectively an iso-singlet and an iso-triplet under a custodial $SU(2)$ -- are relevant with a mass below the cut-off $\Lambda \approx 4\pi v$. 
We study the amplitudes for the processes $W_L W_L \rightarrow W_L W_L, hh, V_{L}V_{L}, V_{L}h$. Requiring unitarity for the elastic channel $W_L W_L \rightarrow W_L W_L$ we can reduce the parameter space to two free parameters for given masses of the heavy vector and the scalar. We study the numerical total cross sections for the associated productions $pp\to Vhjj$ by Vector Boson Fusion (VBF) and $pp\to Vh$ by Drell-Yan (DY) annihilation as functions of these free parameters. The expected rates of same-sign di-lepton and tri-lepton events from the decay of the $Vh$ final state are also given.
\end{abstract}
\vskip 1cm \hspace{0.7cm} May 2010
\end{titlepage}
\setcounter{page}{2}

\tableofcontents

\section{Introduction}
The LHC is finally on and in the next few years it will probably answer many of the open questions in the high energy particle physics. The high c.o.m. energy of the LHC makes it sensitive to energy scales that are
above the Fermi scale $v\approx 246$ GeV, making it possible to discover the mechanism that generates the Electro-Weak Symmetry Breaking (EWSB). The evidence for new phenomena at the Fermi scale, to the least the existence of a Higgs boson, is indirectly written in the most impressive results of LEP and the Tevatron: the high precision measurement of masses and couplings of the weak gauge bosons. These results imply that the gauge sector of the Standard Model (SM) has been tested with very high accuracy. The presence in the theory of massive vector particles tells us that they have to be gauge bosons of a spontaneously broken non-abelian gauge theory and therefore a mechanism that realizes the $SU\(2\)_{L}\times U\(1\)_{Y}\rightarrow U\(1\)_{\text{em}}$ symmetry breaking must be present. Moreover, the violation of unitarity in the scattering of longitudinally polarized gauge bosons, if perturbatively treated, makes the existence of new degrees of freedom universally accepted.

There are in principle two possibilities to realize the EWSB: we can have a weakly coupled dynamics, mediated by at least one fundamental scalar that acquires a \textsc{vev} breaking the Electro-Weak (EW) symmetry or a strong dynamics that becomes non-perturbative above the Fermi scale realizing the breaking through some condensate. In the first case we can expect at least one fundamental scalar particle with a mass below a TeV. On the other hand, if the existence of a light Higgs boson allows the perturbative extrapolation of the theory eventually up to the Planck scale, it introduces the hierarchy problem, i.e. the problem of the stability of the scalar masses to radiative corrections. In the second case the dynamics that breaks the EW symmetry is strongly coupled and makes the theory non-perturbative above the Fermi scale. Some examples of this possibility are given by the \textit{composite} Higgs boson models \cite{Kaplan:1983, Chivukula:1993, Contino:2009, Giudice:2007, Low:2009, Zerwekh:2009}, or even by the Higgs-less models in which some new vector particles control unitarity of the longitudinal gauge boson scattering up to a cut-off $\Lambda \approx 4\pi v$ \cite{Bagger,Pelaez:1996,SekharChivukula:2001,Appelquist:2003, Csaki:2003, Barbieri:2008}.

In the case of strong EWSB, the phenomenology for the pair production of new particles has a very important role. In fact, if the single production of new particles can certainly be the first manifestation of these new physics at the LHC \cite{Dobado:2000,Han:2003,Birkedal:2005yg,He:2007ge,Accomando:2008jh,Belyaev:2008yj,Cata:2009iy}, the pair production is crucial to distinguish among the different models since it is sensitive to many couplings and, in some sense, more model dependent. A study of the associated production of a light Higgs and a weak gauge bosons in a model with vector resonances was done for example in \cite{Zerwekh:2005wh}. More recently, a model independent study of the scalar pair production was done in \cite{Contino:2010}, while the corresponding study was made for vectors in \cite{Barbieri:2010} by means of suitable effective Lagrangians. In both of these recent papers it is shown how the requirement of unitarity in the different inelastic channels for the longitudinal gauge boson scattering strongly constrains the parameter space for the relevant couplings. For a reasonable effective theory approach one can only accept relatively small deviations of the parameters from those corresponding to a good asymptotic behavior of the various physical amplitudes, since large deviations quickly lower the cut-off of the theory making it unacceptably small.

In this paper we are interested to the case in which both a vector and a light scalar are relevant with a mass below the cut-off $\Lambda\approx 3$ TeV. In this case the role of unitarization of the different scattering
channels is played both by the scalar and the vector (an example of this phenomenon is discussed for  Technicolor models in \cite{Foadi:2008}). In particular, the unitarity in the elastic longitudinal gauge boson scattering does not completely constrain the couplings of the scalar and the vector to the gauge bosons, but implies a relation among them. Therefore in this case there is a wider region in the parameter space that is reasonable from the point of view of unitarity, at least in the elastic channel. In this framework we are interested to study the modifications to the phenomenology of the scalar and vector pair productions and the phenomenology of the associated scalar-vector production, that is peculiar to the present case\footnote{We shall not impose the constraints coming from the EW Precision Tests since further effects can be present, e.g. due to new fermionic degrees of freedom, that obscure their interpretation and/or a strong sensitivity to the physics at the cut-off may be involved which we do not pretend to control.}.

The organization of the paper is the following. In the Sect. \ref{sec2} we introduce our Lagrangian that describes the spectrum of a theory with a new scalar $h$ and a new vector $V^{a}$ with masses below the cut-off $\Lambda\approx 4\pi v$, respectively a singlet and a triplet under a custodial $SU\(2\)$ symmetry. In Sect. \ref{sec3} we calculate the scattering amplitudes for the two body processes $W_{L}W_{L}\to W_{L}W_{L},~V_{L}V_{L},~hh,~V_{L}h$. In Sect. \ref{sec4} we discuss the asymptotic behavior of these amplitudes and the constraints on the parameter space imposed by unitarity. In Sect. \ref{sec5} we compute the total cross sections for the associated scalar-vector production at the LHC at $14$ TeV for different values of the parameters. The phenomenology of the same sign di-lepton and tri-lepton events for a high integrated luminosity phase of the LHC is described in Sect. \ref{sec6}. Conclusions are discussed in Sect. \ref{sec7}. Finally in the Appendix \ref{app1} we show how our Lagrangian, for special values of the parameters, describes a $SU\(2\)_{L}\times SU\(2\)_{C}\times U\(1\)_{Y}$ gauge theory spontaneously broken by two Higgs doublets.

\section{The basic Lagrangian}\label{sec2}
We are interested to study a scalar-vector system in the framework of Strongly Interacting EWSB by adopting an approach as model independent as possible. Nevertheless, for our approach to make sense at all we have to make some assumptions. One way to state these assumptions is the following:
\begin{enumerate}
\item Before weak gauging, the Lagrangian responsible for EWSB has a $SU\left(2\right)_{L}\times SU\left(2\right)^{N}\times SU\left(2\right)_{R}$ global symmetry, with $SU\left(2\right)^{N}$ gauged, spontaneously broken to the diagonal $SU\left(2\right)_{d}$ by a generic non-linear sigma model;
\item Only one vector triplet $V_{\mu}^{a}$ of the $SU\left(2\right)^{N}$ gauge group has a mass below the cut-off $\Lambda\approx 3\,\text{TeV}$, while all the other heavy vectors can be integrated out. Furthermore the new vector triplet $V_{\mu}^{a}$ couples to fermions only through the mixing with the weak gauge bosons of $SU\(2\)_{L}\times U\(1\)_{Y}$, $Y=T_{3R}+1/2(B-L)$;
\item The spectrum also contains a scalar singlet of $SU\left(2\right)_{d}$ with a relatively low mass $m_{h}\apprle v$. 
\end{enumerate}
We believe that these assumptions may represent a physically interesting situation.

\subsection{A single heavy vector below the cut-off}
Let us consider a gauge theory based on the global symmetry $G=SU\left(2\right)_{L}\times SU\left(2\right)^{N}\times SU\left(2\right)_{R}$ broken down to the diagonal subgroup $H=SU\left(2\right)_{d}$ by a general non-linear sigma model given by
\be\label{nlsm}
\La_{\chi}^{\(N\)}=\sum_{I,J}v_{IJ}^{2}\left<D_{\mu}\Sigma_{IJ}\left(D^{\mu}\Sigma_{IJ}\right)^{\dag}\right>-\sum_{i}\f{1}{2g_{i}^{2}}\left<F_{\mu\nu\,i}F^{\mu\nu}_{i}\right>\,,\qquad \Sigma_{IJ}\to g_{I}\Sigma_{IJ}g_{J}^{\dag}
\ee
where $D_{\mu}$ is the covariant derivative of $G$, $v_{IJ}$ are the breaking scales corresponding to the $\Sigma_{IJ}$ link fields, $F_{\mu\nu}^{i}$ and $g_{i}$ are respectively the field strength tensor and the gauge coupling of the vector $V_{i}$, $g_{I,J}$ are elements of the various $SU\left(2\right)$ and $\left<\right>$ denotes the $SU\(2\)$ trace. If we are interested to the special case in which in addition to the EW gauge bosons only one vector triplet $V_{\mu}^{a}$ under $H$ has a mass below the cut-off $\Lambda\approx 4\pi v$, the theory can be described using an effective Lagrangian based on the $SU\(2\)_{L}\times SU\(2\)_{R}/SU\(2\)_{L+R}$ non-linear sigma model plus $SU\(2\)_{L}\times SU\(2\)_{R}$ invariant kinetic term and interactions for a massive spin-1 field with suitable couplings, as described in \cite{Barbieri:2010}.
Let us briefly recall this construction.

We start from the kinetic Lagrangian for the $SU\(2\)_{L}\times SU\(2\)_{R}/SU\(2\)_{L+R}$ non-linear field $U$ and for the EW gauge bosons\footnote{Our normalizations are such that $M_{W}=\f{gv}{2}$ with  $v\approx 246$ GeV.}
\be\label{lbasic}
\La_{\chi}=\f{v^{2}}{4}\left<D_{\mu}U\(D^{\mu}U\)^{\dag}\right>-\f{1}{2g^{2}}\left<W_{\mu\nu}W^{\mu\nu}\right>-\f{1}{2g^{\prime 2}}\left<B_{\mu\nu}B^{\mu\nu}\right>\,,
\ee
where
\be
\begin{array}{ll}
\displaystyle D_{\mu}U=\demub U-iB_{\mu}U+iUW_{\mu}\,,\qquad  & U=e^{\f{i\pi}{v}}\,, \qquad \pi=\pi^{a}\tau^{a}\,,\\\\
\displaystyle W_{\mu\nu}=\demub W_{\nu}-\denub W_{\mu}-i[W_{\mu},W_{\nu}]\,,\qquad & W_{\mu}=\f{g}{2}W_{\mu}^{a}\tau^{a}\,,\\\\
\displaystyle B_{\mu\nu}=\demub B_{\nu}-\denub B_{\mu}\,, & B_{\mu}=\f{g'}{2}B_{\mu}^{0}\tau^{3}\,,\\\\
\end{array}
\ee
and $\tau^{a}$ are the usual Pauli matrices. Now let us consider a new massive vector $V_{\mu}=\f{1}{\sqrt{2}}V_{\mu}^{a}\tau^{a}$ belonging to the adjoint representation of $SU\(2\)_{L+R}$. The $SU\(2\)_{L}\times SU\(2\)_{R}$-invariant kinetic Lagrangian for $V_{\mu}$ is the usual Proca Lagrangian for a massive vector
\be
\La^{V}_{\text{kin}}=-\f{1}{4}\left<\hat{V}_{\mu\nu}\hat{V}^{\mu\nu}\right>+\f{M_{V}^{2}}{2}\left<V_{\mu}V^{\mu}\right>\,,
\ee
where the field strength tensor $\hat{V}_{\mu\nu}=\nabla_{\mu}V_{\nu}-\nabla_{\nu}V_{\mu}$ is written in terms of the $SU\(2\)_{L}\times SU\(2\)_{R}$ covariant derivative
\be
\nabla_{\mu}V_{\nu}=\demub V_{\nu}+[\Gamma_{\mu},V_{\nu}]\,,
\ee
with the connection $\Gamma_{\mu}$ given by
\be
\Gamma_{\mu}=\f{1}{2}\Big[u^{\dag}\(\demub-iB_{\mu}\)u+u\(\demub-iW_{\mu}\)u^{\dag}\Big]\,,\qquad u\equiv \sqrt{U}\,,\qquad \Gamma_{\mu}^{\dag}=-\Gamma_{\mu}\,.
\ee
To construct the interactions among the new vector and the SM particles we only have another quantity that transforms covariantly with respect to $SU\(2\)_{L+R}$ that is $u_{\mu}=u^{\dag}_{\mu}=iu^{\dag}D_{\mu}U u^{\dag}$. Using these ingredients we can write an interaction Lagrangian for the new vector in such a way that this vector exactly describes the lighter gauge vector (a part from the EW gauge bosons) of the general theory \eqref{nlsm} in the limit in which all the other heavy vectors have a mass above the cut-off $\Lambda\approx 4\pi v$ \cite{Barbieri:2010}. The interaction Lagrangian relevant for our purposes is\footnote{The coefficient of the second term in \eqref{LVint}, i.e. the mixing term between the $V$ and the EW gauge bosons, is usually kept independent on $g_{V}$ and called $f_{V}$. However, since the general gauge invariance of the model \eqref{nlsm} always requires $f_{V}=2g_{V}$ and since we are not interested here in studying deviations from this general gauge theory, we have explicitly eliminated the parameter $f_{V}$ in favor of the parameter $g_{V}$ (see e.g. \cite{Barbieri:2010} and the references cited therein). The last term in \eqref{LVint}, not involving the vector $V$, also arises from \eqref{nlsm} and is crucial to prevent the appearance of contributions growing like $s^{2}$ in the longitudinal $WW\to WW$ scattering amplitude.} 
\be\label{LVint}
\begin{array}{lll}
\displaystyle \La^{V}_{\text{int}}&=&\displaystyle -\f{ig_{V}}{2\sqrt{2}}\left<\hat{V}_{\mu\nu}[u^{\mu},u^{\nu}]\right>-\f{g_{V}}{\sqrt{2}}\left<\hat{V}_{\mu\nu}\(uW^{\mu\nu}u^{\dag}+u^{\dag}B^{\mu\nu}u\)\right>\\\\
&&\displaystyle +\f{ig_{K}}{4\sqrt{2}}\left<\hat{V}_{\mu\nu}[V^{\mu},V^{\nu}]\right>-\f{1}{8}\left<[V_{\mu},V_{\nu}][u^{\mu},u^{\nu}]\right>+\f{g_{V}^{2}}{8}\left<[u_{\mu},u_{\nu}][u^{\mu},u^{\nu}]\right>\,,
\end{array}
\ee 
where $g_{V}$ and $g_{K}$ are linear combinations of the gauge couplings of the different $SU\(2\)$ groups in \eqref{nlsm}. In particular in the case in which instead of the general gauge theory \eqref{nlsm} we consider the gauge theory $SU\(2\)_{L}\times SU\(2\)_{C}\times U\(1\)_{Y}$ spontaneously broken by a non-linear sigma model, these couplings are simply related to the gauge coupling $g_{C}$ of the $SU\(2\)_{C}$ by
\be
g_{K}=\f{1}{g_{V}}=2g_{C}\,.
\ee
Moreover, while in this special case the mass of the vector is related to the couplings by
\be
M_{V}=g_{C}v=\f{g_{K}v}{2}=\f{v}{2g_{V}}\,,
\ee
in the general case of \eqref{nlsm} there are mass mixings between the different gauge vectors and the relations between the mass $M_{V}$ and the couplings $g_{V}$ and $g_{K}$ also depend on the couplings of the other vectors above the cut-off. In the general case the trilinear coupling $g_{K}$ get spread among the mass eigenstates $V^{\mu}_{i}$ so that the trilinear coupling in the Lagrangian becomes
\be
\La_{V^{3}}=\f{i\tilde{g}_{K}^{ijk}}{4\sqrt{2}}\left<\hat{V}_{\mu\nu}^{i}[V^{\mu}_{j},V^{\nu}_{k}]\right>\,.
\ee
It is simple to see that indicating with $\tilde{g}_{V_{i}}$ the mass eigenstates couplings corresponding to $g_{V_{i}}$ and keeping only the lightest vector we have $\tilde{g}_{K}^{111}\tilde{g}_{V_{1}}\neq 1$. Anyway in the general case the sum rule
\be
\sum_{i}\tilde{g}_{V_{i}}\tilde{g}_{K}^{ijj}=1
\ee
can be proved.\\
Summarizing we can describe the theory \eqref{nlsm} in the case of a single heavy vector below the cut-off $\Lambda\approx 4\pi v$ using the effective Lagrangian
\be\label{lvector}
\La^{V}=\La_{\chi}+\La^{V}_{\text{kin}}+\La^{V}_{\text{int}}\,.
\ee

\subsection{Adding a scalar}
As mentioned, we also want to consider the possibility that a scalar particle exists below the cut-off. In principle this light scalar could be a Strongly Interacting Light Higgs (SILH) boson in the sense of \cite{Giudice:2007} or a more complicated object arising from an unknown strong dynamics. The couplings of this particle to the SM particles and to the heavy vector $V$ will be strongly related to the mechanism that generates it. The measurement of the different cross sections that are sensitive to the different couplings, hopefully at the LHC but eventually also at a future Linear Collider, could give information about this mechanism.

Assuming that the light scalar is parity even and is a singlet under $SU\(2\)_{L+R}$, the most general Lagrangian to describe it relevant for our purposes is given by two pieces:
\be\label{lscalar}
\La_{h}=\f{1}{2}\demub h \demua h +\f{m_{h}^{2}}{2}h^{2}+\f{v^{2}}{4}\left<D_{\mu}U\(D^{\mu}U\)^{\dag}\right>\(2a\f{h}{v}+b\f{h^{2}}{v^{2}}\)\,,
\ee
that contains the kinetic term, the mass term and the interactions with the Standard Model particles and
\be\label{lvhinteraction}
\La_{h-V}=\f{dv}{8g_{V}^{2}}h\left<V_{\mu}V^{\mu}\right>
\ee
that represents the interaction with the heavy vector $V$. Here $a$, $b$ and $d$ are dimensionless constants. 
Putting together the Lagrangians \eqref{lvector}, \eqref{lscalar} and \eqref{lvhinteraction}
we obtain the complete effective Lagrangian
\be\label{ltot}
\La_{\text{eff}}=\La^{V}+\La_{h}+\La_{h-V}\,.
\ee
We show in Appendix \ref{app1} that the Lagrangian \eqref{ltot}, for the special values
\be\label{gaugeparameters}
a=\f{1}{2}\,,\quad b=\f{1}{4}\,,\quad d=1\,,\quad g_{K}=\f{1}{g_{V}}\,,\quad g_{V}=\f{v}{2M_{V}}\,,
\ee
is obtained from a gauge theory based on $SU\(2\)_{L}\times SU\(2\)_{C}\times U\(1\)_{Y}$ spontaneously broken by two Higgs doublets (with the same \textsc{vev}) in the limit $m_{H}\gg \Lambda$ for the mass of the $L$-$R$-parity odd scalar $H$\footnote{As we discuss in Appendix \ref{app1} the mass of the $L$-$R$-parity odd scalar $H$ can be simply raised above the cut-off without any further hypothesis on the low energy physics.}.

\section{Two body $W_{L}W_{L}$ scattering amplitudes}\label{sec3}
In this Section we compute the scattering amplitudes:
\be\label{processes}
\begin{array}{lcl}
\displaystyle \Amp\(W^{a}_{L}W^{b}_{L}\to W^{c}_{L}W^{d}_{L}\)\qquad\qquad&&\qquad\qquad \displaystyle \Amp\(\pi^{a}\pi^{b}\to \pi^{c}\pi^{d}\)\\
\displaystyle \Amp\(W^{a}_{L}W^{b}_{L}\to V^{c}_{L}V^{d}_{L}\)&\Longrightarrow&\qquad\qquad \displaystyle -\Amp\(\pi^{a}\pi^{b}\to V^{c}_{L}V^{d}_{L}\)\\
\displaystyle \Amp\(W^{a}_{L}W^{b}_{L}\to hh\) &\sqrt{s} \gg M_{W}&\qquad\qquad \displaystyle -\Amp\(\pi^{a}\pi^{b}\to hh\)\\
\displaystyle \Amp\(W^{a}_{L}W^{b}_{L}\to V^{c}_{L}h\)&&\qquad\qquad \displaystyle -\Amp\(\pi^{a}\pi^{b}\to V^{c}_{L}h\)\,,
\end{array}
\ee
where we make use of the Equivalence Theorem to relate the scattering amplitudes involving the Goldstone bosons with the high energy limit of the scattering amplitudes involving the longitudinal polarization of the weak gauge bosons\footnote{The minus sign in the last three amplitudes in \eqref{processes} is due to the fact that the Equivalence Theorem has a factor $\(-i\)^{N}$ where $N$ is the number of external longitudinal vector bosons.}. To simplify the explicit formulae we take the limit $g'=0$ (that implies $Z\approx W^{3}$) so that the $SU\(2\)_{L+R}$ invariance is preserved by the scattering amplitudes. We can study the four processes one by one.

\begin{itemize}
\item $\pi^{a}\pi^{b}\to \pi^{c}\pi^{d}$ scattering amplitude\\
Using the $SU\(2\)_{L+R}$ invariance and the Bose symmetry the amplitude for the four pion scattering can be written in the form
\be
\Amp\(\pi^{a}\pi^{b}\to \pi^{c}\pi^{d}\)=\Amp\(s,t,u\)^{\pi\pi\to \pi\pi}\delta^{ab}\delta^{cd}+\Amp\(t,s,u\)^{\pi\pi\to \pi\pi}\delta^{ac}\delta^{bd}+\Amp\(u,t,s\)^{\pi\pi\to\pi\pi}\delta^{ad}\delta^{bc}\,.
\ee
It receives contributions from the four pion contact interaction $\pi^{4}$ and from the exchange of $W$, $V$ and $h$. The contribution coming from the exchange of a $W$ boson is sub-leading in the sense of the Equivalence Theorem, i.e. is of order $M_{W}/\sqrt{s}$ and therefore we can write
\be
\Amp\(\pi^{a}\pi^{b}\to \pi^{c}\pi^{d}\)=\Amp\(\pi^{a}\pi^{b}\to \pi^{c}\pi^{d}\)_{\pi^{4}}+\Amp\(\pi^{a}\pi^{b}\to \pi^{c}\pi^{d}\)_{V}+\Amp\(\pi^{a}\pi^{b}\to \pi^{c}\pi^{d}\)_{h}\,,
\ee
so that we obtain
\be\label{pipipipi}
\Amp\(s,t,u\)^{\pi\pi\to \pi\pi}=\f{s}{v^{2}}+\f{g_{V}^{2}M_{V}^{2}}{v^{4}}\Big[-3s+M_{V}^{2}\(\f{\(u-s\)}{t-M_{V}^{2}}+\f{\(t-s\)}{u-M_{V}^{2}}\)\Big]-\f{a^{2}}{v^{2}}\(\f{s^{2}}{s-m_{h}^{2}}\)\,.
\ee

\item $\pi^{a}\pi^{b}\to V^{c}_{L}V^{d}_{L}$ scattering amplitude\\
The amplitude can be reduced to
\be
\Amp\(\pi^{a}\pi^{b}\to V^{c}_{L}V^{d}_{L}\)=\Amp\(s,t,u\)^{\pi\pi\to VV}\delta^{ab}\delta^{cd}+\mathcal{B}\(s,t,u\)^{\pi\pi\to VV}\delta^{ac}\delta^{bd}+\mathcal{B}\(s,u,t\)^{\pi\pi\to VV}\delta^{ad}\delta^{bc}\,.
\ee
It receives contributions from the $\pi^{2}V^{2}$ contact interaction and the exchange of $\pi$, $V$ and $h$
\be\label{vvcontributions}
\begin{array}{lll}
\displaystyle \Amp\(\pi^{a}\pi^{b}\to V_{L}V_{L}\)&=&\displaystyle \Amp\(\pi^{a}\pi^{b}\to V_{L}V_{L}\)_{\pi^{2}V^{2}}+\Amp\(\pi^{a}\pi^{b}\to V_{L}V_{L}\)_{\pi}\\
&& \displaystyle+ \Amp\(\pi^{a}\pi^{b}\to V_{L}V_{L}\)_{V}+\Amp\(\pi^{a}\pi^{b}\to V_{L}V_{L}\)_{h}\,.
\end{array}
\ee
The explicit forms obtained for $\Amp\(s,t,u\)^{\pi\pi\to V_{L}V_{L}}$ and $\mathcal{B}\(s,t,u\)^{\pi\pi\to V_{L}V_{L}}$ are
\begin{align}
& \displaystyle \Amp\(s,t,u\)^{\pi\pi\to V_{L}V_{L}}=  \f{g_{V}^{2}M_{V}^{2}s}{v^{4}\(s-4M_{V}^{2}\)}\Big[\f{\(t+M_{V}^{2}\)^{2}}{t}+\f{\(u+M_{V}^{2}\)^{2}}{u}\Big]+\f{ad}{2v^{2}}\(\f{s}{s-m_{h}^{2}}\)\(s-2M_{V}^{2}\)\,,\label{pipiVVA}\\
& \displaystyle \mathcal{B}\(s,t,u\)^{\pi\pi\to V_{L}V_{L}} = \f{t-u}{2v^{2}}-\f{g_{V}^{2}M_{V}^{2}s\(u+M_{V}^{2}\)^{2}}{v^{4}u\(s-4M_{V}^{2}\)}+\f{s\(u-t\)}{4v^{2}M_{V}^{2}}\(g_{V}g_{K}\f{s+2M_{V}^{2}}{s-M_{V}^{2}}-1\)\,.\label{pipiVVB}
\end{align}

\item $\pi^{a}\pi^{b}\to hh$ scattering amplitude\\
The amplitude can be written as
\be
\Amp\(\pi^{a}\pi^{b}\to hh \)=\Amp\(s,t,u\)^{\pi\pi\to hh}\delta^{ab}\,.
\ee
This amplitude receives contributions from the $\pi^{2}h^{2}$ contact interaction and the exchange of $\pi$ and $h$
\be\label{hhcontributions}
\Amp\(\pi^{a}\pi^{b}\to hh \)= \Amp\(\pi^{a}\pi^{b}\to hh \)_{\pi^{2}h^{2}}+\Amp\(\pi^{a}\pi^{b}\to hh \)_{\pi}+ \Amp\(\pi^{a}\pi^{b}\to hh \)_{h}\,.
\ee
In this case $\Amp\(s,t,u\)^{\pi\pi\to hh}$ is given by
\be\label{pipihh}
\Amp\(s,t,u\)^{\pi\pi\to hh}= -\f{1}{v^{2}}\(s\(b-a^{2}\)+\f{3as m_{h}^{2}}{2\(s-m_{h}^{2}\)}-2a^{2}m_{h}^{2}+\f{a^{2}m_{h}^{4}}{t}+\f{a^{2}m_{h}^{4}}{u}\)\,.
\ee

\item $\pi^{a}\pi^{b}\to V^{c}_{L}h$ scattering amplitude\\
The $SU\(2\)_{L+R}$ invariance implies
\be
\Amp\(\pi^{a}\pi^{b}\to V^{c}_{L}h \)=\Amp\(s,t,u\)^{\pi\pi\to Vh}\epsilon^{abc}\,.
\ee
The amplitude receives contributions from the exchange of $\pi$ and $V$
\be
\Amp\(\pi^{a}\pi^{b}\to V^{c}_{L}h \)= \Amp\(\pi^{a}\pi^{b}\to V^{c}_{L}h \)_{\pi}+ \Amp\(\pi^{a}\pi^{b}\to V^{c}_{L}h \)_{V}
\ee
so that the explicit value of $\Amp\(s,t,u\)^{\pi\pi\to Vh}$ is
\be\label{pipiVh}
\begin{array}{lll}
\Amp\(s,t,u\)^{\pi\pi\to Vh}
&=&\displaystyle \f{i\(t-u\)}{2v\sqrt{\(M_{V}^{2}+m_{h}^{2}-s\)^{2}-4m_{h}^{2}M_{V}^{2}}}\Bigg[\f{d}{4g_{V}M_{V}}\f{s}{s-M_{V}^{2}}\(m_{h}^{2}-M_{V}^{2}-s\)\\
&&\displaystyle +\f{2ag_{V}M_{V}}{v^{2}tu}\Big[m_{h}^{2}M_{V}^{2}\(m_{h}^{2}-M_{V}^{2}+s\)+tu\(M_{V}^{2}-m_{h}^{2}+s\)\Big]\Bigg]\,.
\end{array}
\ee
\end{itemize}

\section{Asymptotic amplitudes and parameter constraints}\label{sec4}
In the very high energy limit in which $s\gg M_{V}^{2}\gg m_{h}^{2}$ we can summarize the amplitudes \eqref{pipipipi}, \eqref{pipiVVA}, \eqref{pipiVVB}, \eqref{pipihh} and \eqref{pipiVh} as follows:
\begin{subequations}\label{asymptamplitudes}
\begin{align}
& \displaystyle \Amp\(s,t,u\)^{\pi\pi\to\pi\pi} \approx \f{s}{v^{2}}\(1-a^{2}-\f{3g_{V}^{2}M_{V}^{2}}{v^{2}}\)+\f{g_{V}^{2}M_{V}^{4}}{v^{4}}\Big[\(\f{\(u-s\)}{t}+\f{\(t-s\)}{u}\)\Big]\label{asymptamplitudes1}\,,\\
& \displaystyle \Amp\(s,t,u\)^{\pi\pi\to VV} \approx \(\f{ad}{2v^{2}}-\f{1}{4v^{2}}\)\(s-2M_{V}^{2}\)\label{asymptamplitudes2}\,,\\
& \hspace{-1.5mm}\begin{array}{lll} \displaystyle \mathcal{B}\(s,t,u\)^{\pi\pi\to VV} &\approx&\displaystyle \f{u-t}{2v^{2}}\bigg[\f{s}{2M_{V}^{2}}\(g_{V}g_{K}-1\)-1+\f{3g_{V}g_{K}}{2}\(1+\f{M_{V}^{2}}{s}\)\bigg]\vspace{2mm}\\
&&\displaystyle -\f{g_{V}^{2}M_{V}^{2}u}{v^{4}}\(1+\f{4M_{V}^{2}}{s}+\f{2M_{V}^{2}}{u}\)\,,\end{array}\label{asymptamplitudes3}\\
& \displaystyle \Amp\(s,t,u\)^{\pi\pi\to hh} \approx -\f{1}{v^{2}}\Big[\(b-a^{2}\)s+\f{a m_{h}^{2}}{2}\(3-4a\)\Big]\label{asymptamplitudes4}\,,\\
& \hspace{-1.5mm}\begin{array}{lll} \displaystyle \Amp\(s,t,u\)^{\pi\pi\to Vh} &\approx&\displaystyle \f{ig_{V}M_{V}\(t-u\)}{v}\Bigg[\f{a}{v^{2}}-\f{d}{8g_{V}^{2}M_{V}^{2}}\Bigg]\vspace{2mm}\\
&&\displaystyle+\f{ig_{V}M_{V}\(t-u\)}{vs}\Bigg[\f{a}{v^{2}}\(M_{V}^{2}-m_{h}^{2}\)+\f{d}{8g_{V}^{2}M_{V}^{2}}\(m_{h}^{2}-2M_{V}^{2}\)\Bigg]\,.\end{array}\label{asymptamplitudes5}
\end{align}
\end{subequations}
For generic values of the parameters, all these amplitudes grow with the c.o.m. energy like $s$ except $\mathcal{B}\(s,t,u\)^{\pi\pi\to VV}$ that grows like $s^{2}$. On the other hand, with the parameters as in \eqref{gaugeparameters} the amplitudes reduce to
\begin{subequations}\label{asymptamplitudesgauge}
\begin{align}
& \displaystyle \Amp\(s,t,u\)^{\pi\pi\to\pi\pi} \approx \f{M_{V}^{2}}{4v^{2}}\Big[\(\f{\(u-s\)}{t}+\f{\(t-s\)}{u}\)\Big]+O\(\f{m_{h}^{2}}{v^{2}}\)\label{asymptamplitudesgauge1}\,,\\
& \displaystyle \Amp\(s,t,u\)^{\pi\pi\to VV} \approx O\(\f{m_{h}^{2}}{v^{2}}\)\label{asymptamplitudesgauge2}\,,\\
& \displaystyle \mathcal{B}\(s,t,u\)^{\pi\pi\to VV} \approx -\f{t}{4v^{2}}-\f{M_{V}^{2}}{4v^{2}}\(\f{u+3t}{s}+2\)\label{asymptamplitudesgauge3}\,,\\
& \displaystyle \Amp\(s,t,u\)^{\pi\pi\to hh} \approx -\f{m_{h}^{2}}{4v^{2}}\label{asymptamplitudesgauge4}\,,\\
& \displaystyle \Amp\(s,t,u\)^{\pi\pi\to Vh} \approx \f{iM_{V}^{2}\(u-t\)}{4v^{2}s}+O\(\f{m_{h}^{2}}{v^{2}}\)\label{asymptamplitudesgauge5}\,.
\end{align}
\end{subequations}
From the last relations we see that with the choice \eqref{gaugeparameters} of the parameters, that corresponds to the choice of the $SU\(2\)_{L}\times SU\(2\)_{C}\times U\(1\)_{Y}$ gauge model spontaneously broken by two Higgs doublets in the limit of very heavy $L$-$R$-parity odd scalar $H$, all the amplitudes except for $\mathcal{B}\(s,t,u\)^{\pi\pi\to VV}$ have a constant asymptotic behavior. As shown in the Appendix \ref{app1} if we add to the spectrum also the $H$ scalar we can regulate also the $\mathcal{B}\(s,t,u\)^{\pi\pi\to VV}$ amplitude making the theory asymptotically well behaved and perturbative.

The choice of parameters as in \eqref{gaugeparameters} is however too restrictive. Other than $g_{V}g_{K}=1$, so that $\pi\pi \to VV$ grows at most like $s$, we only pretend that the exchange of the scalar and of the vector lead together to a good asymptotic behavior of elastic $W_{L}W_{L}$ scattering, i.e. 
\be\label{unitarityrelation}
a=\sqrt{1-\f{3G_{V}^{2}}{v^{2}}}\,,\qquad\qquad G_{V}\equiv g_{V}M_{V}\,.
\ee

The processes \eqref{processes} are all important at the LHC in order to understand the underlying mechanism that can generate the spectrum that we consider. In fact the pair production of new states can be very useful to measure the different couplings and to constrain the parameter space. Both the scalar and vector pair productions have been recently studied in \cite{Contino:2010} and \cite{Barbieri:2010} respectively. The phenomenology studied in that works changes as follows in the present approach:
\begin{itemize} 
\item Scalar pair production\\
Equations \eqref{hhcontributions} shows that there aren't contributions of the heavy vector to the scalar pair production so that the results of \cite{Contino:2010} exactly hold also in this case\footnote{The existence of an $hVV$ vertex can in principle modify the width of the scalar. However it is reasonable to consider negligible this effect.}. 
\item Vector pair production\\
From equation \eqref{vvcontributions} we see that there is a contribution to the heavy vectors pair production coming from the scalar exchange. Unfortunately this contribution is not big enough to compensate the decrease of the cross section due to the lowering of $G_{V}$. In other words, we see that the longitudinal $WW\to WW$ scattering unitarity relation \eqref{unitarityrelation} implies $G_{V}\leq v/\sqrt{3}$. This region of values of $G_{V}$ is rather below the value considered in \cite{Barbieri:2010} that is $G_{V}=200$~GeV. This effect leads to a fast decrease of the total cross sections that quickly fall out of the LHC accessible region. 
\end{itemize}

It remains to study the associated $Vh$ production that is obviously absent from both the phenomenological studies cited above. The associated production can be generated both by Vector Boson Fusion (VBF) and by Drell-Yan (DY) $q\bar{q}$ annihilation. The gluon-gluon fusion associated production at order $\alpha_{S}^{2}$ could be another relevant production channel, i.e. comparable to the VBF or the DY productions. However, in addition to the loop factor suppression, the absence of a direct coupling of the vector to the quarks (or at least to the top) introduces a further suppression coming from the $WV$ mixing with respect to the analogous case for the Higgs pair production by gluon-gluon fusion in the SM.  There can be relevant two-loop contributions of order $\alpha^2_S$ to the total cross sections but their estimation could become a very difficult task which is beyond the scope of this work. For this reason we defer for a future work the study of the top quark effects in the Vh associated production.
In the next section we discuss the total cross sections for the associated production by VBF and by DY.

\section{Associated production of $Vh$ total cross sections}\label{sec5}
In this section we discuss the total cross section for the associated $Vh$ production of the heavy vector and the light scalar. There are three possible final states for the associated production, corresponding to the three charge states of the $V$: $hV^{-}$, $hV^{0}$ and $hV^{+}$. According to the constraints discussed in the previous Section on the parameter space we can compute the total cross sections for some reference values of the independent parameters, that we choose to be $G_{V}$ and $d$. Some values of the total cross sections at the LHC for $\sqrt{s}=14$ TeV for different values of the parameters and for a scalar mass $m_{h}=180$ GeV are listed in Tables \ref{table1}, \ref{table2} and \ref{table3} for the production of $hV^{-}$, $hV^{0}$ and $hV^{+}$ respectively. 
\begin{table}[htb!]
\begin{minipage}[b]{8.2cm}
\centering
\resizebox{6cm}{!} {
\begin{tabular}[c]{|c|c|c|c|c|}
		\hline
		$G_{V}$ &  $a$ &  $d$ & VBF (fb) & DY (fb) \\
		\hline
		$\sqrt{5} v/4$  &$1/4$ & $0$ & $0.05$& $0$ \\
		\hline
		$\sqrt{5} v/4$  &$1/4$ & $1$ & $0.09$ & $3.31$ \\
		\hline
		$\sqrt{5} v/4$  &$1/4$ & $2$ & $0.62$& $13.24$ \\
		\hline
		$v/2$  & $1/2$ & $0$ & $0.15$ & $0$ \\
		\hline
		$v/2$  & $1/2$ & $1$ & $0.05$ & $4.14$ \\
		\hline
		$v/2$  & $1/2$ & $2$ & $0.56$ & $16.56$ \\
		\hline
		$v/\sqrt{6}$  & $1/\sqrt{2}$ & $0$ & $0.20$  & $0$ \\
		\hline
		$v/\sqrt{6}$  & $1/\sqrt{2}$ & $1$ & $0.08$  & $6.20$ \\
		\hline
		$v/\sqrt{6}$  & $1/\sqrt{2}$ & $2$ & $0.89$  & $24.80$ \\
		\hline
\end{tabular}}
\\\vspace{4mm} \footnotesize{(\ref*{table1}.a)}
 \end{minipage}
\ \hspace{2mm} \hspace{3mm} \ 
\begin{minipage}[b]{8.2cm}
\centering
\resizebox{6cm}{!} {
\begin{tabular}[c]{|c|c|c|c|c|}
		\hline
		$G_{V}$ &  $a$ &  $d$ & VBF (fb) & DY (fb) \\
		\hline
		$\sqrt{5} v/4$  &$1/4$ & $0$ & $0.02$ & $0$ \\
		\hline
		$\sqrt{5} v/4$  &$1/4$ & $1$ & $0.08$ & $1.23$ \\
		\hline
		$\sqrt{5} v/4$  &$1/4$ & $2$ & $0.49$ & $4.92$ \\
		\hline
		$v/2$  & $1/2$ & $0$ & $0.07$ & $0$ \\
		\hline
		$v/2$  & $1/2$ & $1$ & $0.06$ & $1.54$ \\
		\hline
		$v/2$  & $1/2$ & $2$ & $0.48$ & $6.16$ \\
		\hline
		$v/\sqrt{6}$  & $1/\sqrt{2}$ & $0$ & $0.09$  & $0$ \\
		\hline
		$v/\sqrt{6}$  & $1/\sqrt{2}$ & $1$ & $0.09$  & $2.30$ \\
		\hline
		$v/\sqrt{6}$  & $1/\sqrt{2}$ & $2$ & $0.75$  & $9.20$ \\
		\hline
\end{tabular}}
\\\vspace{4mm} \footnotesize{(\ref*{table1}.b)}
 \end{minipage}
\vspace{-7mm}\caption{Total cross sections for the associated production of $hV^{-}$ final state by VBF and DY at the LHC for $\sqrt{s}=14$ TeV as functions of the different parameters for $M_{V}=700$ GeV (\ref*{table1}.a) and $M_{V}=1$ TeV (\ref*{table1}.b). The parameter $a$ is fixed by the value of $G_{V}$ (and vice versa) according to equation \eqref{unitarityrelation}.}\label{table1}
\end{table}
We have chosen $m_{h}=180$ GeV to maximize both the total cross sections and the branching ratio for $h\to W^{+}W^{-}$. In this case signals of the associated productions can appear in the multi-lepton channels. In particular if the final state contains at least a pair of equal sign $W$'s there can be signals in the same-sign di-lepton and tri-lepton final states from W decays that are much simpler to be separated from the background than those ones corresponding to the hadronic final states.
\begin{table}[htb!]
\begin{minipage}[b]{8.2cm}
\centering
\resizebox{6cm}{!} {
\begin{tabular}[c]{|c|c|c|c|c|}
		\hline
		$G_{V}$  &$a$ &  $d$ & VBF(fb) & DY(fb) \\
		\hline
		$\sqrt{5} v/4$ & $1/4$ & $0$ & $0.08$& $0$ \\
		\hline
		$\sqrt{5} v/4$ & $1/4$ & $1$ & $0.14$ & $6.14$ \\
		\hline
		$\sqrt{5} v/4$ & $1/4$ & $2$ & $0.99$& $24.56$ \\
		\hline
		$v/2$  & $1/2$ & $0$ & $0.24$  & $0$ \\
		\hline
		$v/2$ & $1/2$ & $1$ & $0.08$ & $7.67$ \\
		\hline
		$v/2$ & $1/2$ & $2$ & $0.90$  & $30.68$ \\
		\hline
		$v/\sqrt{6}$ & $1/\sqrt{2}$ & $0$ & $0.32$  & $0$ \\
		\hline
		$v/\sqrt{6}$ & $1/\sqrt{2}$ & $1$ & $0.13$ & $11.51$ \\
		\hline
		$v/\sqrt{6}$ & $1/\sqrt{2}$ & $2$ & $1.42$ & $46.04$ \\
		\hline
\end{tabular}}
\\\vspace{4mm} \footnotesize{(\ref*{table2}.a)}
 \end{minipage}
\ \hspace{2mm} \hspace{3mm} \ 
\begin{minipage}[b]{8.2cm}
\centering
\resizebox{6cm}{!} {
\begin{tabular}[c]{|c|c|c|c|c|}
		\hline
		$G_{V}$  &$a$ &  $d$ & VBF(fb) & DY(fb) \\
		\hline
		$\sqrt{5} v/4$ & $1/4$ & $0$ & $0.04$ & $0$ \\
		\hline
		$\sqrt{5} v/4$ & $1/4$ & $1$ & $0.13$ & $2.43$ \\
		\hline
		$\sqrt{5} v/4$ & $1/4$ & $2$ & $0.79$ & $9.74$ \\
		\hline
		$v/2$  & $1/2$ & $0$ & $0.11$  & $0$ \\
		\hline
		$v/2$ & $1/2$ & $1$ & $0.09$ & $3.04$ \\
		\hline
		$v/2$ & $1/2$ & $2$ & $0.78$ & $12.16$ \\
		\hline
		$v/\sqrt{6}$ & $1/\sqrt{2}$ & $0$ & $0.15$  & $0$ \\
		\hline
		$v/\sqrt{6}$ & $1/\sqrt{2}$ & $1$ & $0.15$ & $4.57$ \\
		\hline
		$v/\sqrt{6}$ & $1/\sqrt{2}$ & $2$ & $1.22$ & $18.28$ \\
		\hline
\end{tabular}}
\\\vspace{4mm} \footnotesize{(\ref*{table2}.b)}
 \end{minipage}
\vspace{-8mm}\caption{Total cross sections for the associated production of $hV^{0}$ final state by VBF and DY at the LHC for $\sqrt{s}=14$ TeV as functions of the different constants for $M_{V}=700$ GeV (\ref*{table2}.a) and $M_{V}=1$ TeV (\ref*{table2}.b). The parameter $a$ is fixed by the value of $G_{V}$ (and vice versa) according to equation \eqref{unitarityrelation}.}\label{table2}
\end{table}
Obviously different values of $m_{h}$ are possible: in that case the detection of a signal can be disfavored by the large branching ratio for $h\to b\bar{b}$ for $m_{h}<2M_{W}$, by the large branching ratio for $h\to ZZ$ for $m_{h}>2M_{Z}$ and by the small cross sections for $m_{h}\apprge 250$ GeV (see Fig. \ref{fig2}).
\begin{table}[htb!]
\begin{minipage}[b]{8.2cm}
\centering
\resizebox{6cm}{!} {
\begin{tabular}[c]{|c|c|c|c|c|}
		\hline
		$G_{V}$  & $a$ &  $d$ & VBF(fb) & DY(fb) \\
		\hline
		$\sqrt{5} v/4$ & $1/4$ & $0$ & $0.10$ & $0$ \\
		\hline
		$\sqrt{5} v/4$ & $1/4$ & $1$ & $0.18$  & $7.30$ \\
		\hline
		$\sqrt{5} v/4$ & $1/4$ & $2$ & $1.28$ & $29.20$ \\
		\hline
		$v/2$ & $1/2$ & $0$ & $0.33$  & $0$ \\
		\hline
		$v/2$ & $1/2$ & $1$ & $0.10$  & $9.12$ \\
		\hline
		$v/2$ & $1/2$ & $2$ & $1.15$  & $36.48$ \\
		\hline
		$v/\sqrt{6}$ & $1/\sqrt{2}$ & $0$ & $0.43$  & $0$ \\
		\hline
		$v/\sqrt{6}$ & $1/\sqrt{2}$ & $1$ & $0.17$  & $13.68$ \\
		\hline
		$v/\sqrt{6}$ & $1/\sqrt{2}$ & $2$ & $1.82$  & $54.72$ \\
		\hline
\end{tabular}}
\\\vspace{4mm} \footnotesize{(\ref*{table3}.a)}
 \end{minipage}
\ \hspace{2mm} \hspace{3mm} \ 
\begin{minipage}[b]{8.2cm}
\centering
\resizebox{6cm}{!} {
\begin{tabular}[c]{|c|c|c|c|c|}
		\hline
		$G_{V}$  & $a$ &  $d$ & VBF(fb) & DY(fb) \\
		\hline
		$\sqrt{5} v/4$ & $1/4$ & $0$ & $0.05$ & $0$ \\
		\hline
		$\sqrt{5} v/4$ & $1/4$ & $1$ & $0.18$  & $3.03$ \\
		\hline
		$\sqrt{5} v/4$ & $1/4$ & $2$ & $1.10$ & $12.12$ \\
		\hline
		$v/2$ & $1/2$ & $0$ & $0.16$  & $0$ \\
		\hline
		$v/2$ & $1/2$ & $1$ & $0.12$  & $3.79$ \\
		\hline
		$v/2$ & $1/2$ & $2$ & $1.07$  & $15.16$ \\
		\hline
		$v/\sqrt{6}$ & $1/\sqrt{2}$ & $0$ & $0.22$  & $0$ \\
		\hline
		$v/\sqrt{6}$ & $1/\sqrt{2}$ & $1$ & $0.20$  & $5.69$ \\
		\hline
		$v/\sqrt{6}$ & $1/\sqrt{2}$ & $2$ & $1.66$  & $22.76$ \\
		\hline
\end{tabular}}
\\\vspace{4mm} \footnotesize{(\ref*{table3}.b)}
 \end{minipage}
\vspace{-8mm}\caption{Total cross sections for the associated production of $hV^{+}$ final state by VBF and DY at the LHC for $\sqrt{s}=14$ TeV as functions of the different constants for $M_{V}=700$ GeV (\ref*{table3}.a) and $M_{V}=1$ TeV (\ref*{table3}.b). The parameter $a$ is fixed by the value of $G_{V}$ (and vice versa) according to equation \eqref{unitarityrelation}.}\label{table3}
\end{table}

The total cross sections have been computed using the Matrix Element Generator CalcHEP \cite{calchep} with the CTEQ5M NLO parton distribution functions and the model was implemented in it using the FeynRules Mathematica package \cite{feynrules}. For the calculation of the VBF total cross sections the acceptance cuts $p_{T\,j}>30$ GeV and $|\eta|<5$ for the forward quark jets have been imposed. From the tables we immediately see that the DY total cross sections are much greater than the corresponding VBF ones.
This is due in part to the structure of the phase space, that for the DY is a $2\to2$ and for the VBF is a $2\to4$ and in part to the structure of the squared amplitude that for the DY is proportional to
\be
|\Amp\(q\bar{q}\to Vh\)|^{2}\propto g_{V}^{2}\f{d^{2}}{g_{V}^{4}}=\f{d^{2}}{g_{V}^{2}}\,,
\ee
while the VBF has a more complicated structure that has a strong dependance on $d-a$. 

One important result that emerges from the tables is that if the VBF total cross sections are too small to expect a signal at the LHC, the DY ones can give rise to a signal for a large region of the parameter space. In the next section we give the expected rates of multi-lepton events coming from the total cross sections listed in Tables \ref{table1}-\ref{table3}.

\begin{figure}[!htb]
\centering
\vspace{5mm}\includegraphics[width=13cm]{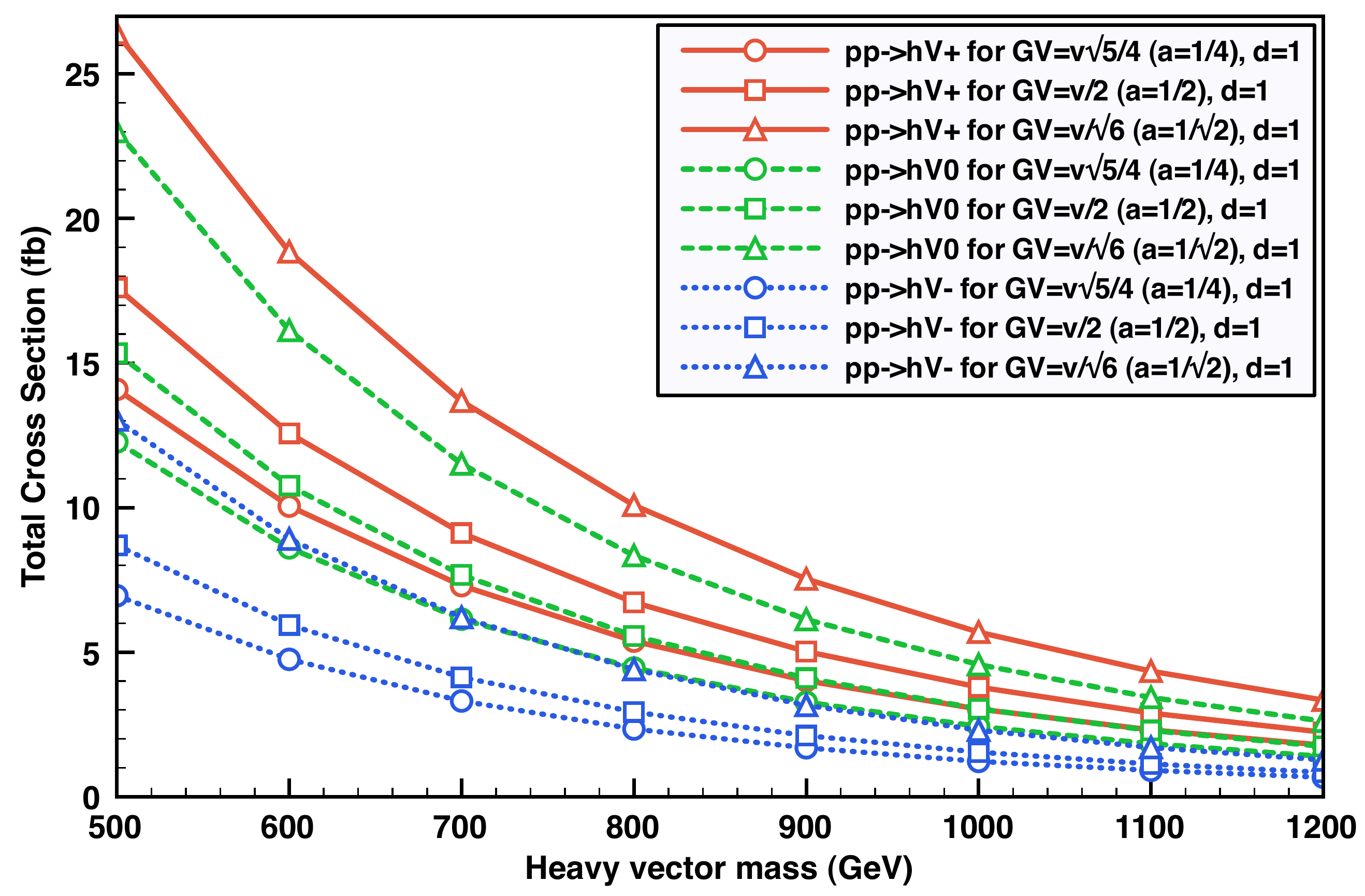}
\vspace{-4mm}\caption{Total cross sections for the $Vh$ associated productions via Drell--Yan $q\bar{q}$ annihilation as functions of the heavy vector mass at the LHC for $\sqrt{s}=14$ TeV, for $m_{h}=180$ GeV, for different values of $G_{V}$ (corresponding to different values of $a$ according to \eqref{unitarityrelation}) and for $d=1$. Since the DY total cross sections are proportional to $d^{2}$ the results can be simply generalized to different values of $d$.}
\label{fig1}
\end{figure}
In Figure \ref{fig1} the total cross sections for the DY associated production at the LHC for $\sqrt{s}=14$ TeV as functions of the heavy vector mass for different values of the parameter $G_{V}$ (and therefore of $a$ according to \eqref{unitarityrelation}) are depicted. We see that even for $d=1$ that corresponds to the choice of the gauge model coupling (see App. \ref{app1}) the total cross sections are of order of $10$ fb for a vector mass between $500$ GeV and $800$ GeV. Furthermore, since  the DY total cross sections grow with $d^{2}$, deviations from $d=1$ could result in a strong increase of  the values given in Figure \ref{fig1}.

Finally, to give an idea of the dependence of the total cross sections on the scalar mass $m_{h}$ we plot in Fig. \ref{fig2} the total cross sections for the $Vh$ associated production as functions of the scalar mass for $150\,\text{GeV}<m_{h}<300\, \text{GeV}$. From Fig. \ref{fig2} we immediately see that the total cross sections have almost halved going form $m_h=180$ GeV to $m_h=270$ GeV. Taking also into account the relevant branching ratio of $h$ we can conclude that a scalar with a mass between $2M_{W}$ and $2M_{Z}$ is the most favorable situation to find a signal of the associated production, while it can be much more difficult to access a signal for $m_{h}<2M_{W} $ or $m_{h}>2M_{Z}$.

\begin{figure}[!htb]
\centering
\vspace{5mm}\includegraphics[width=13cm]{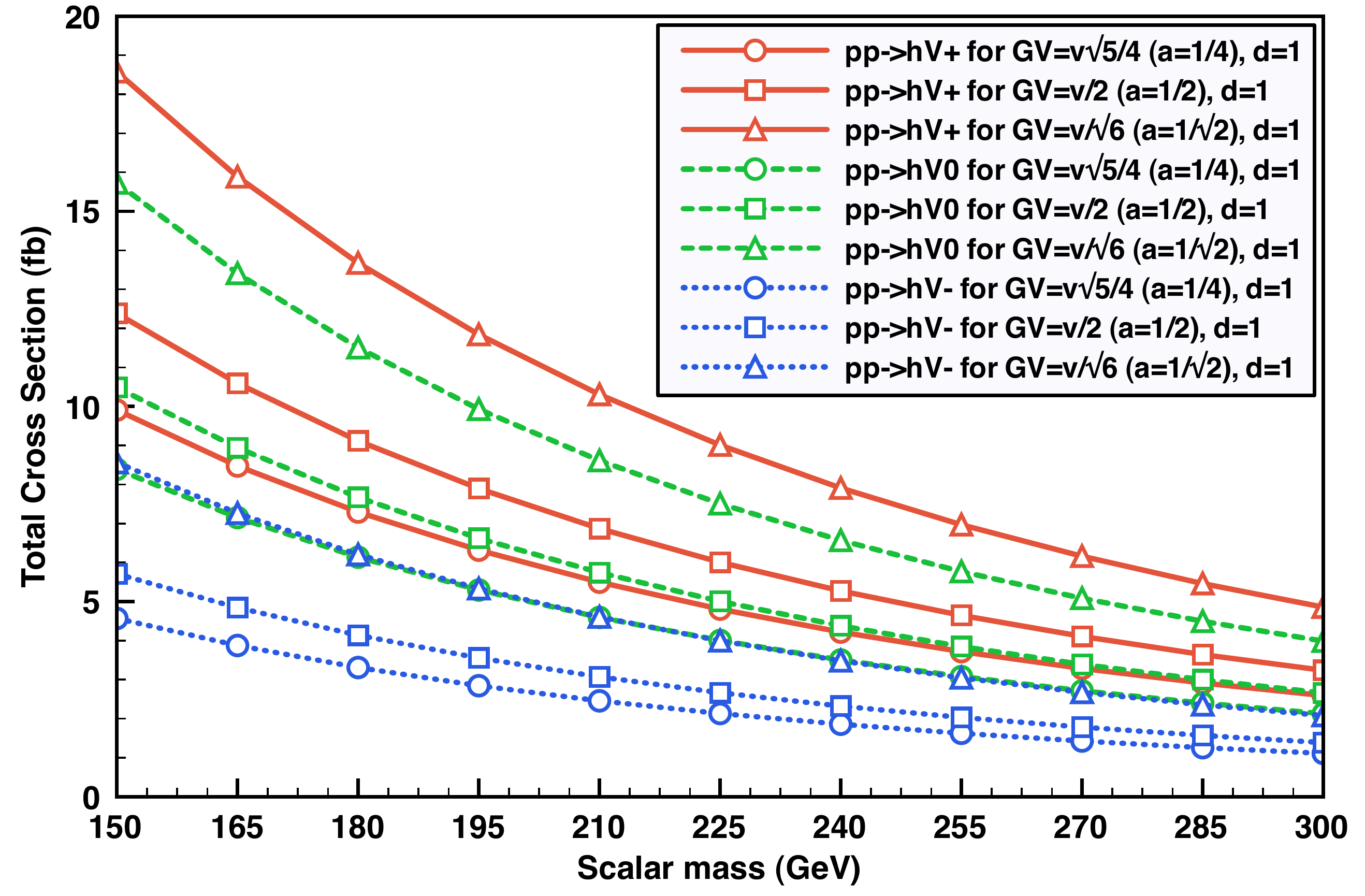}
\vspace{-4mm}\caption{Total cross sections for the $Vh$ associated productions via Drell--Yan $q\bar{q}$ annihilation as functions of the scalar mass at the LHC for $\sqrt{s}=14$ TeV, for $M_{V}=700$ GeV, for different values of $G_{V}$ (corresponding to different values of $a$ according to \eqref{unitarityrelation}) and for $d=1$. Since the DY total cross sections are proportional to $d^{2}$ the results can be simply generalized to different values of $d$.}
\label{fig2}
\end{figure}

\section{Same-sign di-lepton and tri-lepton events}\label{sec6}
The number of multi-lepton events is strongly dependent on the decay modes of the light scalar and the heavy vector. Since the vector couples to the fermions only via the mixing with the weak gauge bosons the width of $V$ into fermions is strongly suppressed with respect to the width into gauge bosons. In the limit $g'\approx 0$ we can write \cite{Barbieri:2008}
\be
\f{\Gamma\(V^{0}\to \bar{\psi}\psi\)}{\Gamma\(V^{0}\to W_{L}^{+}W_{L}^{-}\)}\approx \f{4M_{W}^{4}}{M_{V}^{4}}\,,
\ee
so that we can take the branching ratios
\be
\text{BR}\(V^{+}\to W_{L}^{+}Z_{L}\)\approx \text{BR}\(V^{0}\to W_{L}^{+}W_{L}^{-}\)\approx 1\,.
\ee
For what concerns the scalar $h$ we neglect $\Gamma\(h\to \bar{\psi}\psi\)$ with respect to $\Gamma\(h\to W^{+}W^{-}\)$.

\begin{table}[htb!]
\centering
\small\begin{tabular}[c]{|c|c|c|c|}
		\hline
		Decay Mode & di-leptons ($\%$)& tri-leptons ($\%$)\\
		\hline  
		$V^{0}h\to W^{+}W^{-}W^{+}W^{-}$ & $8.9$ & $3.2$\\
		\hline  
		$V^{\pm}h\to W^{\pm}ZW^{+}W^{-}$ & $4.5$ & $1.0$\\
		\hline  
\end{tabular}\caption{Decay modes and cumulative branching ratios for the different charge configurations of the $hV$ system assuming $BR\(h\to W^{+}W^{-}\)\approx 1$. For the same sign di-lepton and tri-lepton branching ratios we consider only the $e$ and $\mu$ leptons coming from the $W$ decays.}\label{BR}
\end{table}

Using the values of the branching fractions given in Table \ref{BR} and a reference integrated luminosity of $\int\mathcal{L}dt=100~\text{fb}^{-1}$ we obtain the total number of same sign di-lepton and tri-lepton events given in Table \ref{events}.

\begin{table}[htb!]
\small\begin{minipage}[b]{8.2cm}
\centering
\begin{tabular}[c]{|c|c|c|c|}
		\hline
		$G_{V}$ &  $a$ & di-leptons & tri-leptons \\
		\hline
		$\sqrt{5}v/4$  &$1/4$ & $102.4$ & $30.3$ \\
		\hline
		$v/2$  & $1/2$ & $128.0$ & $37.8$ \\
		\hline
		$v/\sqrt{6}$  & $1/\sqrt{2}$ & $192.0$  & $56.7$ \\
		\hline
\end{tabular}
\\\vspace{3mm} \footnotesize{(\ref*{events}.a)}
 \end{minipage}
\ \hspace{2mm} \hspace{3mm} \ 
\begin{minipage}[b]{8.2cm}
\centering
\begin{tabular}[c]{|c|c|c|c|c|}
		\hline
		$G_{V}$ &  $a$ & di-leptons & tri-leptons \\
		\hline
		$\sqrt{5}v/4$  &$1/4$ & $41.0$ & $12.0$ \\
		\hline
		$v/2$  & $1/2$ & $51.0$ & $15.1$ \\
		\hline
		$v/\sqrt{6}$  & $1/\sqrt{2}$ & $76.6$  & $22.6$ \\
		\hline
\end{tabular}
\\\vspace{3mm} \footnotesize{(\ref*{events}.b)}
 \end{minipage}
\caption{Total number of same sign di-lepton and tri-lepton events ($e$ or $\mu$ from $W$ decays) for the DY $Vh$ associated production at the LHC for $\sqrt{s}=14$ TeV and $\int\mathcal{L}dt=100$ fb$^{-1}$ for $M_{V}=700$ GeV (\ref*{events}.a) and $M_{V}=1$ TeV (\ref*{events}.b) for different values of the parameter $G_{V}$ (or $a$ according to equation \eqref{unitarityrelation}) and for $d=1$. Since the DY total cross sections are proportional to $d^{2}$ the results can simply be generalized to different values of $d$.}\label{events}
\end{table}

\section{Summary and conclusions}\label{sec7}

If EWSB is triggered by a strong dynamics, relatively light degrees of freedom may occur which play a special role in preserving unitarity in longitudinal $WW$ scattering. In this work we have considered the case in which such role is played simultaneously by a Higgs-like scalar $h$ and by a vector $V^{a}$ triplet under the custodial $SU\(2\)$. The interactions of these states can be approximately described by an effective Lagrangian invariant under $SU\(2\)_{L}\times SU\(2\)_{R}/SU\(2\)_{L+R}$. Furthermore, for the effective Lagrangian description to make sense at all, we have restricted the interactions of the $V^{a}$ among themselves and with the electroweak gauge bosons to those resulting from a $SU\(2\)_{L}\times SU\(2\)^{N}\times SU\(2\)_{R}$ gauge theory spontaneously broken to the diagonal $SU\(2\)_{d}$ subgroup by a generic non-linear sigma model .

In this framework we have computed the two body amplitudes for the scattering of the $W_{L}W_{L}$ initial state into the $W_L W_L, hh, V_{L}V_{L}, V_{L}h$ final states in terms of $5$ couplings ($a$, $b$, $d$, $g_{V}$ and $g_{K}$) and the two masses $m_{h}$ and $M_{V}$. The relation of these amplitudes with those arising in an explicit $SU\(2\)_{L}\times SU\(2\)_{C}\times U\(1\)_{Y}$ gauge model spontaneously broken by Higgs multiplets has been clarified. The parameter space has been restricted by requiring unitarity in the elastic $W_{L}W_{L}\to W_{L}W_{L}$ channel.

From a phenomenological point of view we have studied the associated production of a scalar and a heavy vector by VBF and DY annihilation. We have found that for a vector with a mass between $500$ GeV and $1$ TeV and for $m_{h}=180$ GeV, the main production mechanism at LHC at $\sqrt{s}=14$ TeV is by DY annihilation. The order of magnitude of the cross sections is about $10$ fb for a reasonable choice of the parameters. This value can also be strongly increased since it depends quadratically on the scalar-vector coupling $d$. The expected same sign di-lepton and tri-lepton events are of order of $10-100$ for an integrated luminosity of $100$ fb$^{-1}$. To see if these events can be made to emerge from the background requires a careful study that is beyond the scope of this work.

\appendix
\section{A well behaved theory at all energies}\label{app1}
Let us consider the following $SU(2)_{L}\times SU(2)_{C}\times U(1)_{Y}$ invariant non-linear sigma model Lagrangian: 
\begin{equation}\label{la0}
\mathcal{L}^{\text{gauge}}=\mathcal{L}_{\chi}^{\text{gauge}}-\frac{1}{2g_{C}^{2}}\left< v_{\mu\nu}v^{\mu\nu}\right> -\frac{1}{2g^{2}}\left<  W_{\mu\nu}W^{\mu\nu}\right>  -\frac{1}{2g^{\prime 2}}\left<  B_{\mu\nu}B^{\mu\nu}\right>  -V\left(  \Sigma_{YC},\Sigma_{CL}\right)\,,
\end{equation} 
where 
\begin{equation}\label{l2a}
v_{\mu}=\frac{g_{C}}{2}v_{\mu}^{a}\tau^{a}
\end{equation} 
is the $SU(2)_{C}$-gauge vector, 
\begin{equation}\label{l3}
\mathcal{L}_{\chi}^{\text{gauge}}=\frac{v^{2}}{2}\left\langle D_{\mu}\Sigma_{YC}\left(  D^{\mu}\Sigma_{YC}\right)  ^{\dag}\right\rangle +\frac{v^{2}}{2}\left\langle D_{\mu}\Sigma_{CL}\left(  D^{\mu}\Sigma_{CL}\right)  ^{\dag}\right\rangle
\end{equation} 
is the symmetry breaking Lagrangian and $V\left(  \Sigma_{YC},\Sigma_{CL}\right)  $ is the scalar potential which has the form
\begin{align} \label{l4}
V\left(  \Sigma_{YC},\Sigma_{CL}\right)   &  =\f{\mu^{2}v^{2}}{2}\left\langle \Sigma_{YC}\Sigma_{YC}^{\dag}\right\rangle +\f{\mu^{2}v^{2}}{2}\left\langle \Sigma_{CL}\Sigma_{CL}^{\dag}\right\rangle -\f{\lambda v^{4}}{4}\left(  \left\langle\Sigma_{YC}\Sigma_{YC}^{\dag}\right\rangle \right)^{2}\nonumber\\ 
&  -\f{\lambda v^{4}}{4}\left(  \left\langle \Sigma_{CL}\Sigma_{CL}^{\dag}\right\rangle \right)  ^{2}-\kappa v^{4}\left\langle \Sigma_{YC}\Sigma_{CL}^{\dag}\Sigma_{CL}\Sigma_{YC}^{\dag}\right\rangle\,.
\end{align} 
To ensure the correct normalization for the Goldstone bosons kinetic terms, $\Sigma_{YC}$ and $\Sigma_{CL}$ are defined as: 
\begin{equation}\label{s17}
\Sigma_{YC}=\left(  1+\frac{h+H}{2v}\right)  U_{YC}\,,\hspace{2cm}\hspace{2cm}U_{YC}=\exp\left[  \frac{i}{2v}\left(  \pi+\sigma \right)\right] \,,
\end{equation}
\begin{equation}\label{s18}
\Sigma_{CL}=\left(  1+\frac{h-H}{2v}\right)  U_{CL}\,,\hspace{2cm}\hspace{2cm}U_{CL}=\exp\left[  \frac{i}{2v}\left(  \pi-\sigma\right)\right]  \,,
\end{equation} 
where $\pi=\pi^{a}\tau^{a}$ and $\sigma=\sigma^{a}\tau^{a}$, being $\pi^{a}$ and $\sigma^{a}$ the Goldstone bosons associated with the EW gauge bosons $W_{\mu}^{a}$ and with the heavy vectors $v_{\mu}^{a}$ respectively and $\tau^{a}$ the usual Pauli matrices. Furthermore $h$ and $H$ are the physical $L$-$R$-parity even and odd scalars respectively, are assumed to have the same \textsc{vev} $v$ and have masses
\be\label{s17b}
m_{h}^{2}=4v^{2}\(\lambda+\kappa\)\,,\qquad\qquad m_{H}^{2}=4v^{2}\(\lambda-\kappa\)\,.
\ee
The two Higgs doublets realize the spontaneous breaking of the $SU(2)_{L}\times SU(2)_{C}\times U(1)_{Y}$ local symmetry to $U(1)_{\text{em}}$, while the global group $G=SU(2)_{L}\times SU(2)_{C}\times SU(2)_{R}$ is broken to the diagonal subgroup $H=SU(2)_{L+C+R}$. The covariant derivatives appearing in \eqref{l3} are given by
\begin{equation}\label{l5a}
D_{\mu}U_{YC}=\partial_{\mu}U_{YC}-iB_{\mu}U_{YC}+iU_{YC}v_{\mu}\,,\hspace{2cm}D_{\mu}U_{CL}=\partial_{\mu}U_{CL}-iv_{\mu}U_{CL}+iU_{CL}W_{\mu}\,.
\end{equation} 
The $U$ fields can be written as $U_{YC}=\sigma_{Y}\sigma_{C}^{\dag}$ and $U_{CL}=\sigma_{C}\sigma_{L}^{\dag}$ where the $\sigma_{L,C,Y}$ are elements of $SU\left(  2\right)_{L,C,R}/H$ respectively\footnote{Remember that only the generator $T^{3}$ of $SU\(2\)_{R}$ is gauged.}. These $\sigma_{I}$ with $I=L,C,Y$ transform under the full $SU\left(2\right)  _{L}\times SU\left(2\right)  _{C}\times U\left(  1\right)  _{Y}$ as $\sigma_{I}\rightarrow g_{I}\sigma_{I}h^{\dag}$. By applying the gauge 
transformation
\begin{equation}\label{x6}
v_{\mu}^{I}\rightarrow\sigma_{I}^{\dag}v_{\mu}^{I}\sigma_{I}+i\sigma_{I}^{\dag}\partial_{\mu}\sigma_{I}=\Omega_{\mu}^{I},\hspace{2cm}\hspace{2cm}U_{IJ}\rightarrow\sigma_{I}^{\dag}U_{IJ}\sigma_{J}=1\,,
\end{equation} 
the symmetry breaking Lagrangian takes the form
\begin{equation}\label{l6a}
\mathcal{L}_{\chi}^{\text{gauge}}=\frac{v^{2}}{2}\left(  1+\frac{h+H}{2v}\right)^{2}\left\langle \left(  \Omega_{\mu}^{Y}-\Omega_{\mu}^{C}\right)  ^{2}\right\rangle +\frac{v^{2}}{2}\left(  1+\frac{h-H}{2v}\right)  ^{2}\left\langle \left(  \Omega_{\mu}^{L}-\Omega_{\mu}^{C}\right)^{2}\right\rangle \,.
\end{equation} 
After the gauge fixing $\sigma_{Y}=\sigma_{L}^{\dag}=u^{2}=U=e^{\f{i\hat{\pi}}{v}}$ and $\sigma_{C}=1$, which implies that $U_{YC}=U_{CL}$ (i.e. $\hat{\sigma}=0$) and which corresponds to the unitary gauge in which we get ride off the Goldstone bosons associated with the heavy vectors $v_{\mu}^{a}$, the Lagrangian of the previous expression becomes
\begin{equation}\label{lt}
\mathcal{L}_{\chi}^{\text{gauge}}=v^{2}\left(  1+\frac{h^{2}+H^{2}}{4v^{2}}+\frac{h}{v}\right)\left(  \left\langle \left(  v_{\mu}-i\Gamma_{\mu}\right)  ^{2}\right\rangle+\frac{1}{4}\left\langle u_{\mu}u^{\mu}\right\rangle \right)  -\frac{1}{2}\left(  2vH+hH\right)  \left\langle u^{\mu}\left(  v_{\mu}-i\Gamma_{\mu}\right)  \right\rangle\,,
\end{equation} 
where
\begin{equation}\label{x10}
u_{\mu}=\Omega_{\mu}^{Y}-\Omega_{\mu}^{L}=iu^{\dag}D_{\mu}U u^{\dag},\hspace{2cm}\Gamma_{\mu}=\frac{1}{2i}\left(  \Omega_{\mu}^{Y}+\Omega_{\mu}^{L}\right)=\f{1}{2}\Big[u^{\dag}\(\demub -iB_{\mu}\)u+u\(\demub-iW_{\mu}\)u^{\dag}\Big]\,.
\end{equation} 
Now by setting
\begin{equation}\label{x11}
v_{\mu}=V_{\mu}+i\Gamma_{\mu}\,,
\end{equation} 
by using the identity \cite{Ecker:1989yg}
\begin{equation}\label{l2}
v_{\mu\nu}=V_{\mu\nu}-i\left[  V_{\mu},V_{\nu}\right]  +\frac{i}{4}\left[u_{\mu},u_{\nu}\right]  +\frac{1}{2}f_{\mu\nu}^{+}\,,
\end{equation} 
where $f_{\mu\nu}^{+}=uW_{\mu\nu}u^{\dagger}+u^{\dagger}B_{\mu\nu}u$
and by redefining $V_{\mu}\rightarrow\frac{g_{C}}{\sqrt{2}}V_{\mu}$, we obtain the following effective Lagrangian
\begin{equation}\label{s20}
\mathcal{L}^{\text{gauge}}=\mathcal{L}_{h=H=0}+\mathcal{L}_{h,H}\,,
\end{equation} 
where
\begin{align}\label{l9}
\mathcal{L}_{h=H=0} &  = -\frac{1}{2g^{2}}\left\langle W_{\mu\nu}W^{\mu\nu}\right\rangle -\frac{1}{2g^{\prime2}}\left\langle B_{\mu\nu}B^{\mu\nu}\right\rangle -\frac{1}{4}\left\langle V_{\mu\nu}V^{\mu\nu}\right\rangle+\frac{v^{2}}{4}\left\langle D_{\mu}U\(D^{\mu}U\)^{\dag}\right\rangle  +\frac{M_{V}^{2}}{2}\left\langle V_{\mu}V^{\mu}\right\rangle \nonumber\\ 
&  +\frac{ig_{C}}{2\sqrt{2}}\left\langle V_{\mu\nu}\left[  V^{\mu},V^{\nu}\right]  \right\rangle  -\frac{g_{C}^{2}}{8}\left<  \left[  V_{\mu},V_{\nu}\right]  \left[V^{\mu},V^{\nu}\right]  \right>-\frac{i}{4\sqrt{2}g_{C}}\left\langle V_{\mu\nu}\left[  u^{\mu},u^{\nu}\right]  \right\rangle  \nonumber\\ 
&  -\frac{1}{8}\left\langle \left[  V_{\mu},V_{\nu}\right]  \left[  u^{\mu},u^{\nu}\right]  \right\rangle -\frac{1}{2\sqrt{2}g_{C}}\left\langle V_{\mu\nu}f^{+\mu\nu}  \right\rangle  +\frac{i}{4}\left\langle \left[  V^{\mu},V^{\nu}\right]  f^{+\mu\nu}  \right\rangle  \nonumber\\ 
&  +\frac{1}{32g_{C}^{2}}\left\langle \left[  u_{\mu},u_{\nu}\right]  \left[  u^{\mu},u^{\nu}\right]  \right\rangle-\frac{1}{8g_{C}^{2}}\left\langle f_{\mu\nu}^{+} f^{+\mu\nu}  \right\rangle -\frac{i}{8g_{C}^{2}}\left\langle \left[  u^{\mu},u^{\nu}\right]  f^{+\mu\nu}  \right\rangle\,,
\end{align}
\begin{align}\label{s22}
\mathcal{L}_{h,H} &  =\frac{1}{4}\left( h^{2}+H^{2}+4vh\right)\left(  \frac{g_{C}^{2}}{2}\left\langle V_{\mu}V^{\mu}\right\rangle +\frac{1}{4}\left\langle D_{\mu}U\(D^{\mu}U\)^{\dag}\right\rangle \right)   \nonumber\\ 
&  -\frac{g_{C}}{2\sqrt{2}}\left(  2vH+hH\right)  \left\langle u^{\mu}V_{\mu}\right\rangle+\frac{1}{2}\left[  \left(  \partial_{\mu}h\right)  ^{2}+\left(\partial_{\mu}H\right)  ^{2}\right]  -V\left(  h,H\right)\,,
\end{align} 
and with the potential $V\left(h,H\right)$ given by
\begin{align}\label{s23}
V\left(  h,H\right)   &  =-\mu^{2}v^{2}\left(  1+\frac{h+H}{2v}\right)^{2}-\mu^{2}v^{2}\left(  1+\frac{h-H}{2v}\right)  ^{2}+2\kappa v^{4}\left(1+\frac{h+H}{2v}\right)  ^{2}\left(  1+\frac{h-H}{2v}\right)  ^{2}\nonumber\\ 
&  +\lambda v^{4}\left(  1+\frac{h+H}{2v}\right)  ^{4}+\lambda v^{4}\left(1+\frac{h-H}{2v}\right)^{4}\,.
\end{align} 
By taking the mass of the $L$-$R$-parity odd $H$ given in \eqref{s17b} infinitely large (so that it is decoupled from the theory), $\mathcal{L}^{\text{gauge}}$ coincides with $\mathcal{L}_{\text{eff}}$ in \eqref{ltot} up to operators irrelevant for the processes \eqref{processes}, only for the values of the parameters: 
\begin{equation}\label{l5}
\begin{array}{l}
\displaystyle g_{V}=\frac{1}{2g_{C}}=\frac{1}{g_{K}}=\frac{v}{2M_{V}},\hspace{2cm} f_{V}=2g_{V},\hspace{2cm}M_{V}=g_{C}v=\frac{1}{2}g_{K}v=\frac{v}{2g_{V}}\,,\vspace{1mm}\\
\displaystyle a=\frac{1}{2},\hspace{2cm}b=\frac{1}{4},\hspace{2cm}d=1,\hspace{2cm} G_{V}=\frac{v}{2}\,.
\end{array}
\end{equation} 
This implies that when the relations \eqref{l5} are satisfied, $\mathcal{L}_{\text{eff}}$ in \eqref{ltot} reduces to $\mathcal{L}^{\text{gauge}}$ in \eqref{s20} in the limit $m_{H}\gg \Lambda$. Since the theory described by $\mathcal{L}^{\text{gauge}}$ is well behaved at all energies, the relations \eqref{l5} allow to take under control the unitarity of the model that we consider in the paper.

\subsection*{Acknowledgements}
We would like to thank Riccardo Barbieri for many useful suggestions and Enrico Trincherini for suggesting us the main idea of studying the associated production of a light scalar and a heavy vector. We also thank Gennaro Corcella, Slava Rychkov and Riccardo Rattazzi for useful discussions.

\end{document}